\documentclass[conference]{IEEEtran}
\IEEEoverridecommandlockouts
\usepackage{cite}
\usepackage{amsmath,amssymb,amsfonts}
\usepackage{algorithmic}
\usepackage{graphicx}
\usepackage{textcomp}
\usepackage{xcolor}
\def\BibTeX{{\rm B\kern-.05em{\sc i\kern-.025em b}\kern-.08em
    T\kern-.1667em\lower.7ex\hbox{E}\kern-.125emX}}

\usepackage[caption=false,labelformat=simple]{subfig}		

\newtheorem{definitionenv}{Definition}
\newtheorem{lemmaenv}[definitionenv]{Lemma}
\newtheorem{theoremenv}[definitionenv]{Theorem}
\newtheorem{corollaryenv}[definitionenv]{Corollary}
\newtheorem{propositionenv}[definitionenv]{Proposition}
\newtheorem{conjectureenv}[definitionenv]{Conjecture}

\newtheorem{app-lemmaenv}[section]{Lemma}
\newtheorem{remarkenv}[definitionenv]{Remark}

\newenvironment{definition}{\begin{definitionenv}\rm}{\end{definitionenv}}
\newenvironment{lemma}{\begin{lemmaenv}\rm}{\end{lemmaenv}}
\newenvironment{theorem}{\begin{theoremenv}\rm}{\end{theoremenv}}
\newenvironment{corollary}{\begin{corollaryenv}\rm}{\end{corollaryenv}}
\newenvironment{proposition}{\begin{propositionenv}\rm}{\end{propositionenv}}
\newenvironment{conjecture}{\begin{conjectureenv}\rm}{\end{conjectureenv}}
\newenvironment{app-lemma}{\begin{app-lemmaenv}\rm}{\end{app-lemmaenv}}
\newenvironment{remark}{\begin{remarkenv}\rm}{\end{remarkenv}}

\newcommand{\bd}{\begin{definition}}
\newcommand{\ed}{\end{definition}}
\newcommand{\edp}{\hspace*{\fill} $\Box$ \end{definition}}
\newcommand{\bl}{\begin{lemma}}
\newcommand{\el}{\end{lemma}}
\newcommand{\elp}{\hspace*{\fill} $\Box$ \end{lemma}}
\newcommand{\bt}{\begin{theorem}}
\newcommand{\et}{\end{theorem}}
\newcommand{\etp}{\hspace*{\fill} $\Box$ \end{theorem}}
\newcommand{\bc}{\begin{corollary}}
\newcommand{\ec}{\end{corollary}}
\newcommand{\ecp}{\hspace*{\fill} $\Box$ \end{corollary}}
\newcommand{\bcj}{\begin{conjecture}}
\newcommand{\ecj}{\end{conjecture}}
\newcommand{\be}{\begin{example}}
\newcommand{\ee}{\end{example}}
\newcommand{\eep}{\hspace*{\fill} $\Box$ \end{example}}
\newcommand{\bp}{\begin{proposition}}
\newcommand{\ep}{\end{proposition}}
\newcommand{\epp}{\hspace*{\fill} $\Box$ \end{proposition}}
\newcommand{\br}{\begin{remark}}
\newcommand{\er}{\end{remark}}
\newcommand{\erp}{\hspace*{\fill} $\Box$ \end{remark}}

\newcommand{\sG}{{\cal G}}

\usepackage[normalem]{ulem}	

\begin{document}


\title{Comparison of 2D topological codes and their decoding performances
\thanks{KYK and CYL were supported by  the Ministry of Science and Technology (MOST) in Taiwan, under Grant MOST110-2628-E-A49-007.}
}


\author{\IEEEauthorblockN{Kao-Yueh~Kuo \,and\, Ching-Yi~Lai}
\IEEEauthorblockA{\textit{Institute of Communications Engineering,} 
\textit{National Yang Ming Chiao Tung University,} 
Hsinchu 300093, Taiwan. \\
 \{kykuo,   cylai\}@nycu.edu.tw}
}

\maketitle

\begin{abstract}
Topological quantum codes are favored because they allow qubit layouts that are suitable for practical implementation.
An $N$-qubit topological code can be decoded by minimum-weight perfect matching (MWPM) with complexity $O(\text{poly}(N))$ if it is of CSS-type.
Recently it is shown that various quantum codes, including non-CSS codes, can be decoded by an adapted belief propagation with memory effects (denoted MBP) with complexity almost linear in $N$.  
In this paper, we show that various two-dimensional topological codes, CSS or non-CSS, regardless of the layout, can be decoded by MBP, 
including color codes and twisted XZZX codes. 
We will comprehensively compare these codes in terms of code efficiency and decoding performance, assuming perfect error syndromes.
\end{abstract}


\section{Introduction} \label{sec:Intro}
%

Quantum codes can be understood as codes over $\text{GF}(4)$~\cite{CRSS98}.
We consider error operators that are tensor products of Pauli matrices
$
I=[\begin{smallmatrix}
	1&0\\0&1
\end{smallmatrix}],
X=[\begin{smallmatrix}
	0&1\\1&0
\end{smallmatrix}],
Z=[\begin{smallmatrix}
	1&0\\0&-1
\end{smallmatrix}],
$ and $
Y=iXZ
$. 
Let $X_j = I^{\otimes (j-1)}\otimes X\otimes I^{\otimes (N-j)}$ and similarly for~$Z_j$.
Then $\{X_j, Z_j\}_{j=1}^N$ together with $iI^{\otimes N}$ generate the $N$-fold Pauli group $\sG_N$.
An abelian subgroup of $\sG_N$ that does not contain $-I^{\otimes N}$ is called a stabilizer group and its elements are stabilizers.
A stabilizer group with $N-K$ independent generators defines an $[[N,K,D]]$ quantum stabilizer code, which encodes $K$  information qubits into $N$ physical qubits, and $D$ is the minimum distance of the code.
The code space is the joint $+1$ eigenspace of the stabilizers, and we can regard the stabilizer group as ``parity checks" for this quantum code. 
For basics of stabilizer codes, please refer to \cite{GotPhD,CRSS98,NC00}.

A Calderbank--Shor--Steane~(CSS) code is a stabilizer code with a set of independent stabilizer generators that are all Pauli $X$s or $Z$s \cite{CS96,Steane96}. They are of interest for fault-tolerant quantum computation (FTQC). 
Moreover, CSS codes can be treated as classical binary codes and  $X$ and $Z$  errors are separately decoded in a suboptimal way.

A topological code is  a stabilizer code, where the composed qubits are placed on a lattice and only local interactions between qubits are required. A stabilizer generator is defined by a plaquette (or called face)  and it operates nontrivially on only the vertices (qubits) of the plaquette.
Thus stabilizer measurements can be locally done, which is a desirable feature for some technologies, such as superconducting qubits.  
Note that only a subset of the stabilizers defined by all the plaquettes are independent. 
 Since stabilizers will be constantly measured in error correction, the time cost for stabilizer measurements affects the decoding complexity. Thus low-weight stabilizer generators are favored. We will consider the average and maximum weight of stabilizer generators that will be measured, denoted by $w_\text{avg}$ and  $w_\text{max}$, respectively.

If a family of $[[N,K,D]]$ codes has a two-dimensional (2D) lattice representation, the code parameters must satisfy the Bravyi--Poulin--Terhal (BPT) bound \cite{BPT10} 
	\begin{equation}\label{eq:BPT}
		N\ge KD^2/c
	\end{equation}
	for a certain $c$.
This number $c$ characterizes the ``efficiency" of the code family and a code family with large $c$ is desirable.

In general, a topological code is degenerate and has many low-weight stabilizers.
The minimum distance does not necessarily characterize its error performance and one has to see the its performance  over a noisy channel using a decoding procedure.  A related notion is  the error {\it threshold} of the combination of a code family and a decoding procedure, which is defined as the physical error rate, below which the logical error rate can be arbitrarily decreased by increasing the lattice size \cite{DKLP02,RHG06,WFSH10,FWH12}.   
Usually a higher threshold is desirable.

 Kitaev  proposed  a family of toric codes  with qubits placed on the surface of a torus  \cite{Kit03}.
 A toric code has two types of stabilizer generators $XXXX$ or $ZZZZ$ operating on the four vertices of a plaquette (c.f. Fig.~\ref{fig:toric_3x3_ori}).
Toric codes satisfy the BPT bound with $c=1$. 
For higher code rate (higher $c$), one may also consider rotated toric codes~\cite{BM07,HFDM12} (c.f. Fig.~\ref{fig:toric_2&4_rot}). 

Since the torus layout may not  be physically implemented, a  2D planar lattice is desirable
and surface codes (c.f. Fig.~\ref{fig:surf_3x3_ori}) and color codes (c.f. Fig.~\ref{fig:color}) are thus introduced~\cite{BK98,BM06}. 
Surface codes have a rectangular layout.
Color codes, with non-rectangular layouts, have higher efficiency than surface codes but they have stabilizer generators of higher weight.

For a higher ratio of $D^2/N$, one may consider non-CSS topological codes. 
A specific type of codes is defined by a weight-four stabilizer generator $XZZX$ operating on each plaquette of a lattice \cite{KDP11,THD12}. 
The smallest XZZX code is the well-known $[[5,1,3]]$ code~\cite{LMPZ96,GotPhD} when described on a twisted torus (see Fig.~\ref{fig:nonCSS_J1_L2}). However, the decoding of an XZZX code may be more complicated since it is non-CSS.

	We will compare the  toric, surface, color, and XZZX codes in this paper in the perspectives of efficiency, stabilizer weight, decoding performance and complexity. 
	We consider decoding by the minimum-weight perfect matching (MWPM) \cite{Edm65} and the refined belief propagation with memory effects (MBP) \cite{KL21}.
	(See \cite{KL20} for some basics of BP decoding of quantum codes.)
 {In particular, MBP achieves a threshold of 17.5\% for many code families, compared to  15.5\%  by MWPM.}
	The results are summarized in Tables~\ref{tbl:2D_codes} and~\ref{tbl:thrd}.

We  review CSS  and non-CSS topological codes in Secs.~\ref{sec:topo} and~\ref{sec:nonCSS_toric}, respectively, 
and compare the codes in Table~\ref{tbl:2D_codes}.
Their decoding performances are discussed in Sec.~\ref{sec:Dec}, and a summary is given in Table~\ref{tbl:thrd}.
Then we conclude in Sec.~\ref{sec:Conclu}.

\section{CSS Topological Codes} \label{sec:topo}


\subsection{Toric codes} \label{sec:toric}

\begin{figure}
	\centering
	\subfloat[\label{fig:toric_3x3_ori}]{\includegraphics[width=0.16\textwidth]{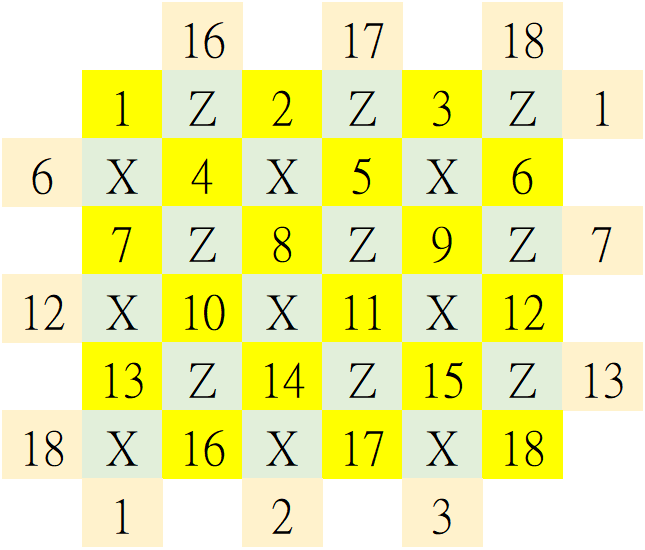}} ~~~~~~~~ 
	\subfloat[\label{fig:surf_3x3_ori}]{\includegraphics[width=0.10\textwidth]{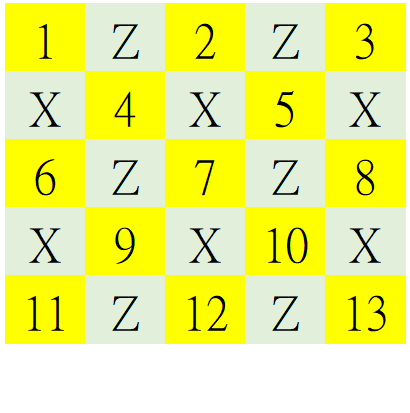}}
	\caption{
		(a) The lattice  of a $[[2L^2, 2, L]]$ toric code for $L=3$.
		(b) The lattice   of  a $[[2L^2-2L+1, 1, L]]$ surface code for $L=3$.
		Each~data qubit is represented by a yellow box labeled by a number, from $1$ to $N$. 
		Each green box with label $W\in\{X,Z\}$ surrounded by four data qubits $i,j,k,l$ represents a stabilizer $W_i W_j W_k W_l$.
		In (a), each orange box on the boundary represents the yellow box with the same label and it is used to  
		indicate the connection of qubits since the lattice is on the surface of a torus.
		For example,  the label $Z$ surrounded by qubits \mbox{1, 2, 4, 16} represents $Z_1 Z_2 Z_4 Z_{16}$.
		%
		In (b), some qubits and stabilizers in (a) are deleted to create physical boundaries.
		Thus there are no wrapped connections and the boundary stabilizers are of weight three, 
		e.g., the label $X$ surrounded by qubits \mbox{1, 4, 6} represents $X_1 X_4 X_6$.
	} \label{fig:toric_surf}		
	\end{figure}

The family of Kitaev's $[[2L^2, 2, L]]$ toric codes has a lattice representation on the surface of a torus \cite{Kit03}, as shown in 
Fig.~\ref{fig:toric_3x3_ori} for $L=3$.
Note that the toric codes have wrapped boundaries (or say, no physical boundaries). 
The toric codes saturate the BPT bound  with efficiency $c=1$.

\begin{figure}
	\centering
	\subfloat[\label{fig:toric_2x2_rot}]{\includegraphics[width=0.08\textwidth]{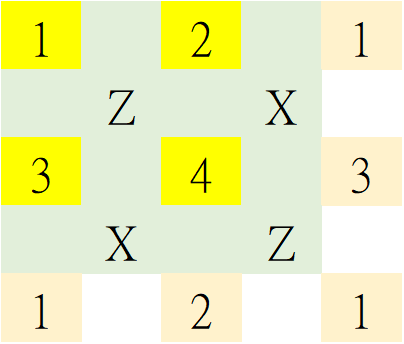}} ~~~~~~~~ 
	\subfloat[\label{fig:toric_4x4_rot}]{\includegraphics[width=0.15\textwidth]{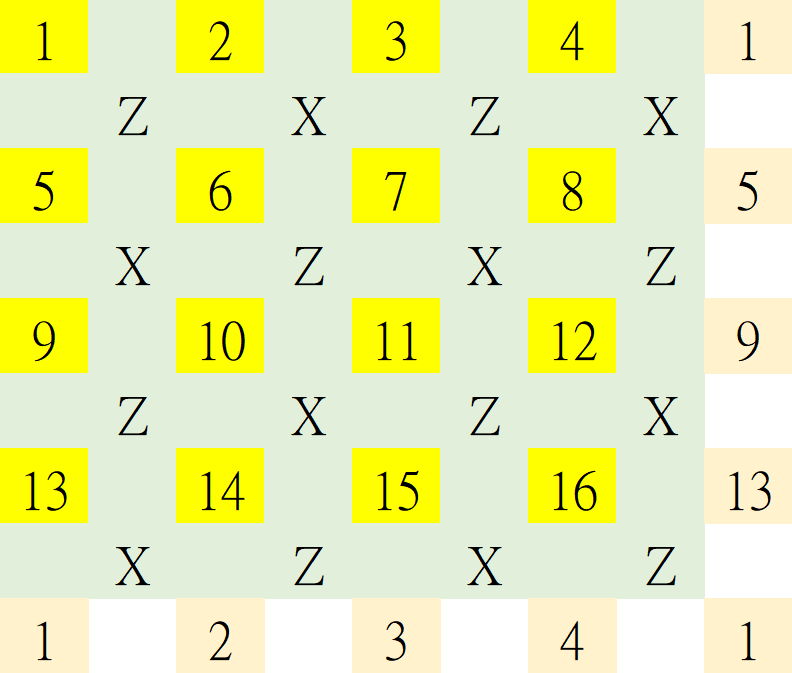}}
	\caption{
		The lattices  of $[[L^2, 2, L]]$ rotated toric codes  for even $L\ge 2$: (a) $L=2$ and (b) $L=4$.
		The labels are similarly defined as  in Fig.~\ref{fig:toric_surf}.
	} \label{fig:toric_2&4_rot}		
	\end{figure}

 A family of $[[L^2, 2, L]]$ \textit{rotated toric codes} with $c=2$ for even $L\ge 2$ has a similar lattice representation \cite{BM07,HFDM12}, as shown in Fig.~\ref{fig:toric_2&4_rot} for $L=2$ and $4$.
 Note that labels $Z$ and $X$ are alternated on each row or column of the lattice. 

Both toric and rotated toric codes have  $w_\text{max}=w_\text{avg}=4$.

\subsection{Surface codes} \label{sec:surf}
The lattice of a toric code is defined on the 2D surface of a torus. 
One can similarly define a lattice on the 2D surface of a plane, by deleting some qubits and stabilizers in a toric code to create physical (or called {\it open}) boundaries \cite{BK98}.
This results in a family of $[[2L^2-2L+1, 1, L]]$ surface codes, as shown in Fig.~\ref{fig:surf_3x3_ori} for $L=3$.
%
%
Since surface codes have a planar lattice, they are more suitable for physical implementation.
However,  the surface codes have lower efficiency $c\approx 1/2$.

\begin{figure}
	\centering 
	\subfloat[\label{fig:surf_3x3_rot}]{\includegraphics[width=0.11\textwidth]{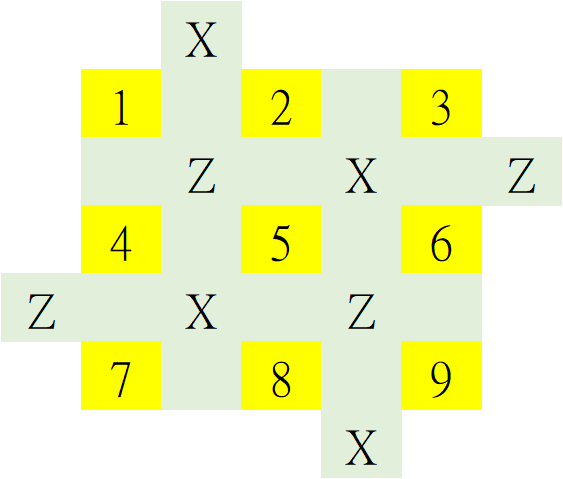}} ~~~~~~~~ 
	\subfloat[\label{fig:surf_5x5_rot}]{\includegraphics[width=0.17\textwidth]{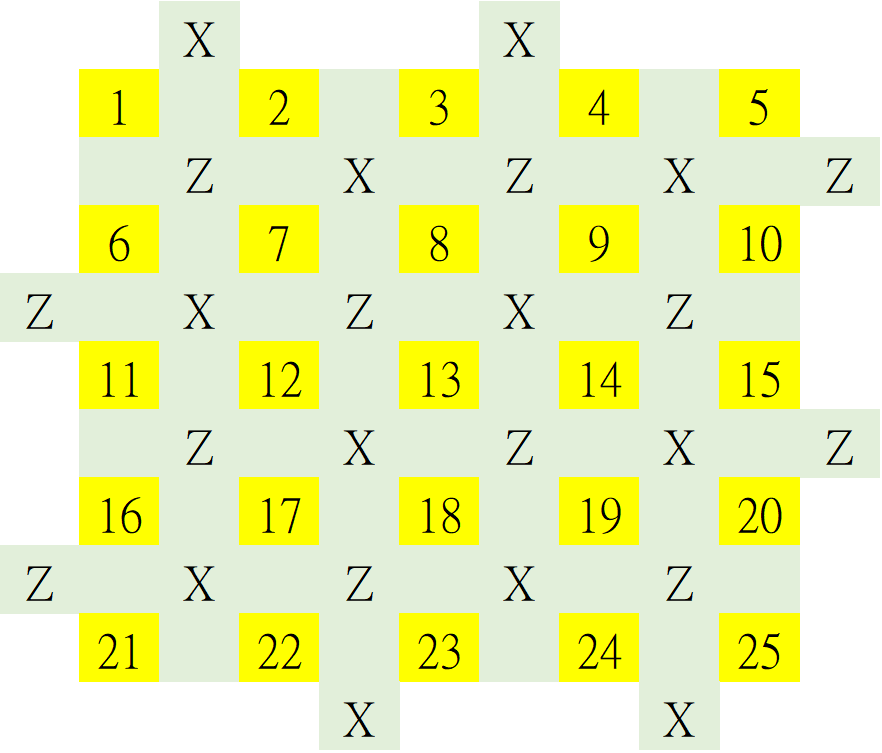}}
	\caption{
		The lattice  of $[[L^2, 1, L]]$ rotated surface codes for odd $L\ge 3$: (a) $L=3$ and (b) $L=5$.
		The labels are similarly defined as  in Fig.~\ref{fig:toric_surf}.
	} \label{fig:surf_3&5_rot}		
	\end{figure}

Similarly, a family of $[[L^2, 1, L]]$ \textit{rotated surface codes} for odd $L\ge 3$  has higher efficiency $c=1$ \cite{BM07,HFDM12}, as shown in Fig.~\ref{fig:surf_3&5_rot} for $L=3$.
%

Both unrotated and rotated surface codes have   $w_\text{max}=4$ and $w_\text{avg}\approx 4$ (asymptotically).

\subsection{Color Codes} \label{sec:color}

\begin{figure}
	\centering 
	\subfloat[\label{fig:color_666}]{\includegraphics[width=0.48\textwidth]{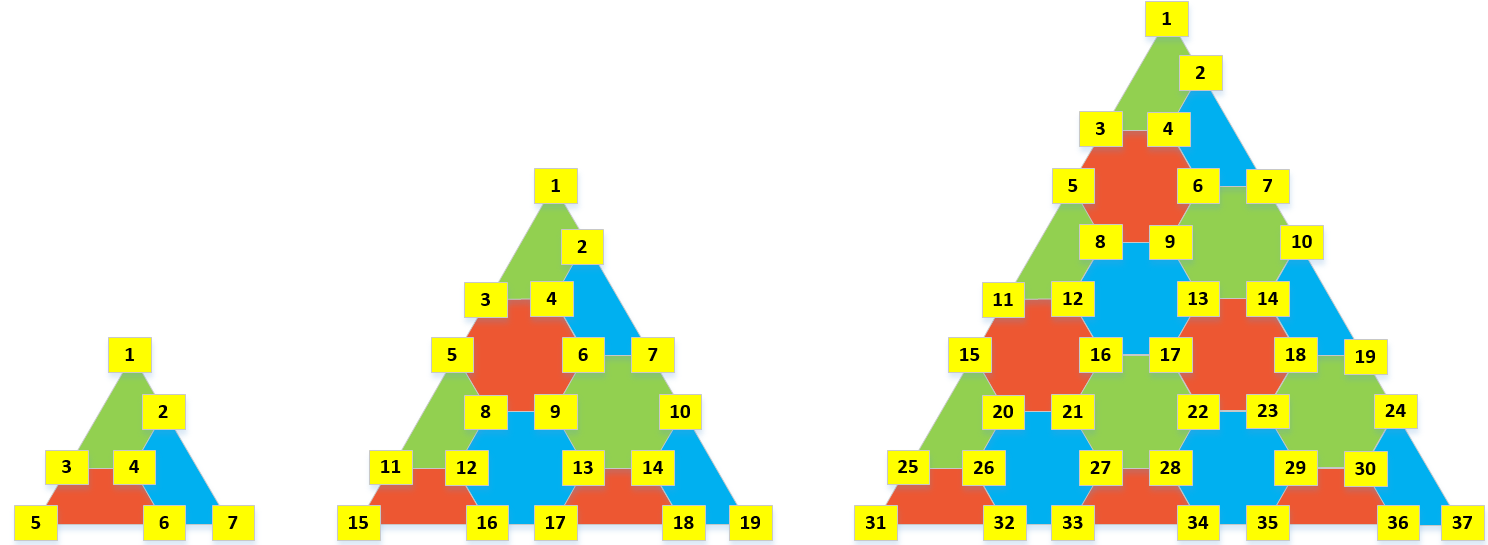}}\\
	\subfloat[\label{fig:color_488}]{\includegraphics[width=0.48\textwidth]{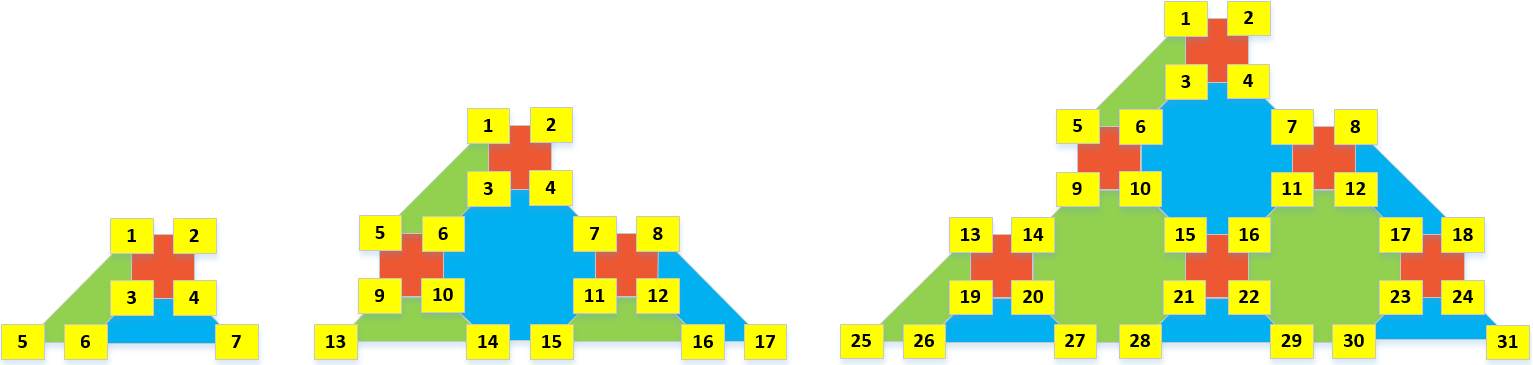}}
	\caption{
		The structures of (a) (6,6,6) color codes and (b) (4,8,8) color codes, for $D=3,5,7$ in each subfigure. 
		There is another typical structure of (4,6,12) color codes, as shown in \cite[Fig.~2(c)]{LAR11}. 
		Each color plaquette defines both an $X$-type stabilizer and a $Z$-type stabilizer. 
		For example, the green plaquette in subfigure (a) defines two stabilizers $X_1X_2X_3X_4$ and $Z_1Z_2Z_3Z_4$.
	} \label{fig:color}		
	\end{figure}

Color codes also have a  planar structure like the surface codes~\cite{BM06}
and three families of color codes are discussed here, as in Fig.~\ref{fig:color} and \cite[Fig.~2(c)]{LAR11}. 
They are designed to have better efficiency $c$ at the cost of larger stabilizer weights.

A color code is   represented by a graph composed of three types of plaquettes (indicated by three colors: red, green, and blue). A data qubit is placed on each vertex in the graph. 
{A plaquette defines both an \mbox{$X$-type} stabilizer and a \mbox{$Z$-type} stabilizer, 
	with an $X$ or $Z$ on each of its vertices.}
There are two rules to build colorful plaquettes.
	First, a plaquette and any of its adjacent plaquettes share two vertices; 
	second, two adjacent plaquettes have different colors. 
The two rules guarantee that the generated stabilizers commute with each other.

The color codes in Fig.~\ref{fig:color} and \cite[Fig.~2(c)]{LAR11} are  {$[[N,1,D]]$ codes with odd $D$}:

\begin{itemize}
\item The (6,6,6) structure has {$N=(3D^2+1)/4$}
	and stabilizer weights $w_\text{max}=6$ and $w_\text{avg}\approx 6$. 

\item The (4,8,8) structure has smaller {$N=(D^2+2D-1)/2$}
	but larger $w_\text{max}=8$ and $w_\text{avg}=(4+8+8)/3 \approx 6.67$. 

\item The (4,6,12) structure has  {$N=(3D^2-6D+5)/2$}
	and larger $w_\text{max}=12$ and $w_\text{avg}=(4+6+12)/3 \approx 7.33$. 
\end{itemize}


\section{Non-CSS Toric Codes} \label{sec:nonCSS_toric}
To have topological codes with 
higher ratio of $D^2/N$,
non-CSS topological codes are studied. 
A plaquette defines \mbox{$X$-type} or \mbox{$Z$-type} stabilizers for CSS topological codes.
Non-CSS stabilizers can be similarly defined by plaquettes \cite{THD12,KDP11} but care needs to be taken for commutation relations. Some general constructions are provided in \cite{SY21}.
Herein, we consider the XZZX codes in \cite{KDP11}, where each code has a rectangular layout rotated by a special angle. 
We provide an interpretation of a twisted torus by combining two regular squared lattices to illustrate the structure of a twisted XZZX code.

\begin{figure}
	\centering
	\subfloat[\label{fig:nonCSS_J0_L2}]{\includegraphics[width=0.08\textwidth]{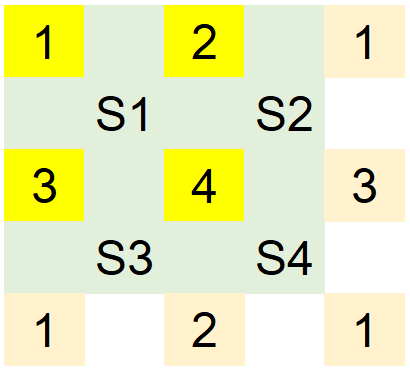}} ~~~~~~~~ 
	\subfloat[\label{fig:nonCSS_J0_L3}]{\includegraphics[width=0.11\textwidth]{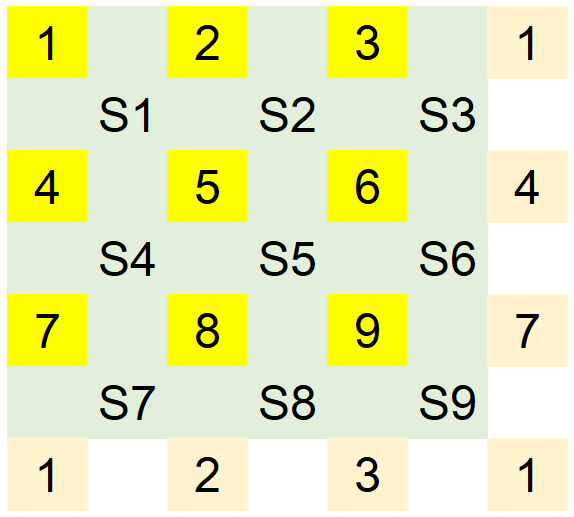}}
	\caption{
		The lattice  of $[[ L^2, 2-(L\%2), L]]$ XZZX toric codes 
		for (a)~$L=2$ and (b)~$L=3$. 
		The  labels are similarly defined as in Fig.~\ref{fig:toric_surf}.
		Label $S$\#  represents a stabilizer $X_iZ_jZ_kX_l$ in the four vertices $i,j,k,l$ on the left-upper, right-upper, left-lower, right-lower corners, respectively.
		For example, $S1$ in (a) is $X_1Z_2Z_3X_4$, and $S9$ in (b) is $X_9Z_7Z_3X_1 = X_1Z_3Z_7X_9$. 
	} \label{fig:nonCSS_J=0}		
	\end{figure}

Recall that a rotated structure has better efficiency.
One can use a lattice like that in Fig.~\ref{fig:toric_2&4_rot} to define a non-CSS code but each plaquette defines a stabilizer of the form $X_iZ_jZ_kX_l$.
This leads to the family of $[[ L^2, 2-(L\%2), L]]$ XZZX toric codes for $L\geq 2$ as shown in Fig.~\ref{fig:nonCSS_J=0}.
This family of codes satisfy the BPT bound with  $c=2$ for even~$L$ and $c=1$ for odd $L$.
Note the XZZX toric codes can be defined for $L\geq 2$ but  rotated toric codes are only defined for even $L$.


 XZZX surface codes can be similarly defined by a lattice structure as in Fig.~\ref{fig:surf_3&5_rot}.

\begin{figure}
	\centering 
	\subfloat[\label{fig:nonCSS_J1_L2}]{\includegraphics[width=0.12\textwidth]{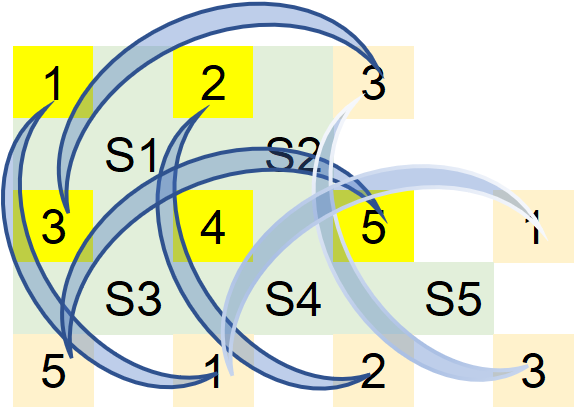}} ~~~~~~~~ 
	\subfloat[\label{fig:nonCSS_J1_L3}]{\includegraphics[width=0.16\textwidth]{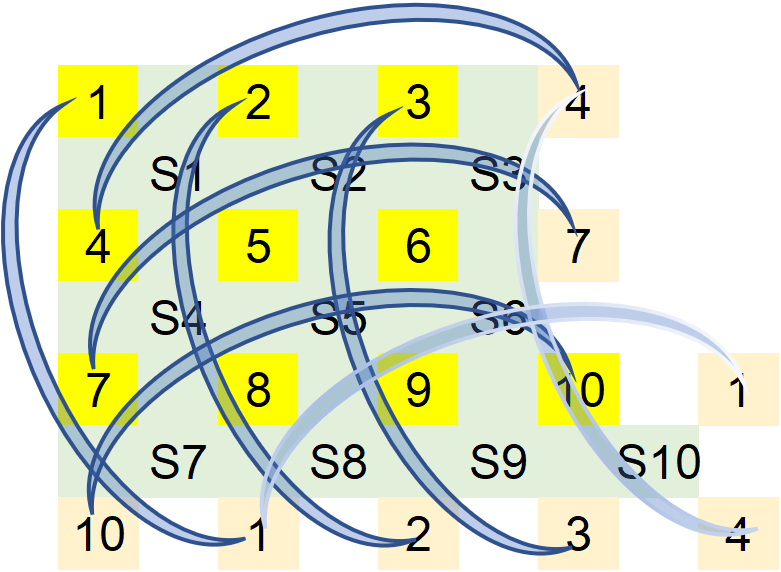}}
	\caption{
		The lattice of twisted XZZX codes with twist $J=1$ for (a)~$L=2$ and (b)~$L=3$. 
		The labels are similarly defined  as in Fig.~\ref{fig:nonCSS_J=0}.
		The (blue) arcs indicate the connections of the wrapped boundaries.
	} \label{fig:nonCSS_J=1}		
	\end{figure}

\begin{figure}
	\centering \centering \includegraphics[width=0.20\textwidth]{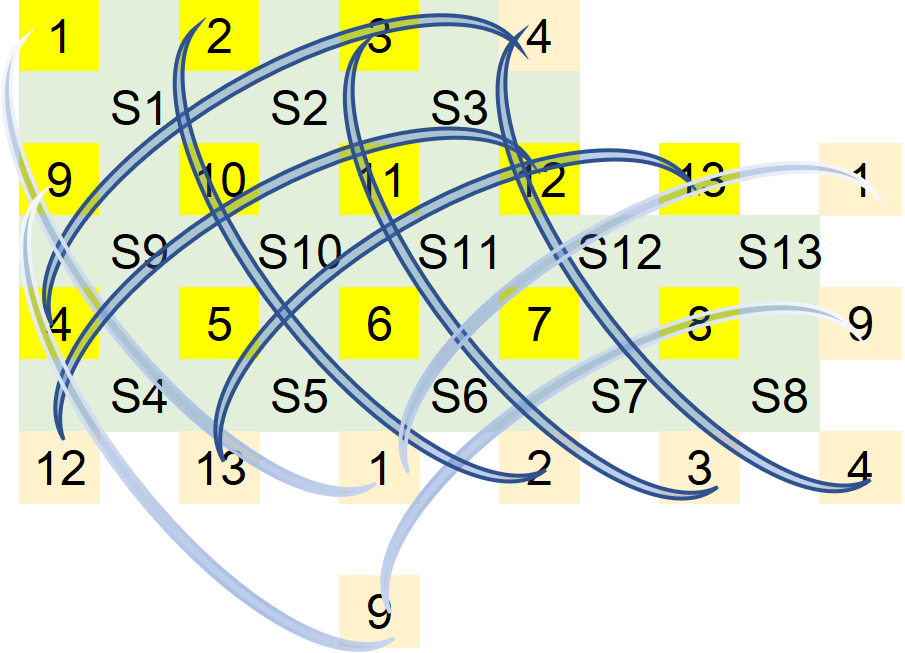}
	\caption{
		The lattice of a twisted XZZX code with  twist $J=2$ for $L=3$, which corresponds to a $[[13, 1, 5]]$ code with stabilizers cyclicly generated by $X_1Z_2Z_9X_{10}$. 
		The   labels are defined similarly as in Fig.~\ref{fig:nonCSS_J=1}.
	} \label{fig:nonCSS_J=2}		
	\end{figure}

To have XZZX codes with higher $D^2/N$, one can adjust the connections of the wrapped boundaries of an  $L\times L$ torus lattice as follows. 
Let $J$ denote an integer twist offset so that the wrapped boundaries have a shift $J$~\cite{Bom10,BBDB13,HG14,YK17,LO18,KPEB18}.
$J$ has to be coprime with $L$, i.e., $\gcd(L,J)=1$,
such that $J$ generates all the elements in $\mathbb Z_L = \{0,1,2,\dots,L-1\}$.
For example, a twist offset $J=1$ in the wrapped boundaries is shown in Fig.~\ref{fig:nonCSS_J=1}.
In addition,  a $J\times J$ lattice needs to be attached to the $L\times L$ lattice so that we have a wrapped structure.
For $J=1$, this is illustrated  in Fig.~\ref{fig:nonCSS_J1_L2} with  qubit 5 and stabilizer $S5$, 
or in Fig.~\ref{fig:nonCSS_J1_L3} with  qubit 10 and stabilizer $S10$.
An example of twist offset $J=2$ is shown in Fig.~\ref{fig:nonCSS_J=2}.

To sum up,  
we have a family of $[[N=L^2+J^2, K, D]]$ twisted XZZX codes, 
where $\gcd(L,J)=1$,  $K=2$, $D=L$ for even $N$, and $K=1$, $D=L+J$ for odd $N$.
We remark that such a twisted XZZX code has its stabilizer generators that can be cyclicly generated.
For example,  the stabilizers in Figs.~\ref{fig:nonCSS_J=1}(a) and (b) can be  cyclicly generated by 
	  $X_1Z_2Z_3X_4$   and $X_1Z_2Z_4X_5$, respectively.
(As a quantum code is like an additive code, the corresponding classical codes over $\text{GF}(4)$ is not necessarily a cyclic linear code.)

\begin{table*}
	\caption{ 			
	Comparison of various codes with 2D topological structures. 
	} 					
	\label{tbl:2D_codes} \centering  
	$\begin{array}{|l|l|l|l|l|l|l|l|}
	\hline
	\text{codes}					& \text{structure}						& [[N,K,D]]			& \text{codes with $D\approx 3$}	& c		& w_\text{avg}	& w_\text{max}	& {\scriptsize\text{planar}}	\\
	\hline
	\text{toric codes \cite{Kit03}}	& \text{Fig.~\ref{fig:toric_3x3_ori}}	& [[2L^2,2,L]]		& [[18,2,3]]	& 1		& 4				& 4				& 				\\
	\text{surface codes \cite{Kit03,BK98}}& \text{Fig.~\ref{fig:surf_3x3_ori}}	& [[2L^2-2L+1,1,L]]	& [[13,1,3]]	& \approx 1/2	& \approx 4		& 4				& \surd			\\
	\text{rotated toric codes  \cite{BM07}}				& \text{Fig.~\ref{fig:toric_2&4_rot} (even $L$)}	& [[L^2,2,L]]	& [[16,2,4]]& 2		& 4		& 4				& 				\\
	\text{rotated surface codes   \cite{BM07}}	& \text{Fig.~\ref{fig:surf_3&5_rot} (odd $L$)}	& [[L^2,1,L]]		& [[9,1,3]]	& 1		& \approx 4		& 4				& \surd			\\
	\text{(6,6,6) color codes \cite{BM06}}		& \text{Fig.~\ref{fig:color_666} (odd $D$)}			& [[\frac{3}{4}D^2+\frac{1}{4},1,D]]	& [[7,1,3]] & \approx 4/3	& \approx 6				& 6	& \surd	\\
	\text{(4,8,8) color codes \cite{BM06}}		& \text{Fig.~\ref{fig:color_488} (odd $D$)}			& [[\frac{1}{2}D^2+D-\frac{1}{2},1,D]]	& [[7,1,3]] & \approx 2		& \approx 6.67	& 8	& \surd	\\
	\text{(4,6,12) color codes \cite{BM06}}		& \text{\cite[Fig.~2(c)]{LAR11} (odd $D$)}			& [[\frac{3}{2}D^2-3D+\frac{5}{2},1,D]]	& [[7,1,3]] & \approx 2/3	& \approx 7.33	& 12& \surd	\\
	\text{XZZX toric codes \cite{THD12}}			& \text{Fig.~\ref{fig:nonCSS_J=0}}	& [[L^2, 2-(L\%2),L]]	& [[9,1,3]], [[16,2,4]] &  2-L\%2	& 4				& 4				& 				\\
	\text{XZZX surface codes \cite{THD12}}		& \text{Fig.~\ref{fig:surf_3&5_rot} {(but plaquette $XZZX$)}} & [[L^2,1,L]] & [[9,1,3]]	& 1		& \approx 4		& 4				& \surd			\\
	\text{XZZX twisted codes \cite{KDP11}}		& \text{Figs.~\ref{fig:nonCSS_J1_L2} and~\ref{fig:nonCSS_J=2} ($J=L-1$)} & [[(D^2+1)/2,1,D]] & [[5,1,3]]	& \approx 2	& 4			& 4				& 				\\
	\hline
	\end{array}$
	%
\end{table*}

For $J=L-1$, we have a family of twisted XZZX codes with parameters $[[(D^2+1)/2,1,D]]$ for $D=2L-1$ for any integer $L\ge 2$.
This family of codes satisfy the BPT bound with efficiency $c\approx 2$.
The smallest nontrivial code in this family is the unique $[[5,1,3]]$ code  in Fig.~\ref{fig:nonCSS_J1_L2}.

Obviously, the twisted XZZX codes  have  $w_\text{max} = w_\text{avg} = 4$.

All the mentioned codes are compared  in Table~\ref{tbl:2D_codes}.
Observe that a toric code family exists with good efficiency $c=2$ and  low stabilizer weights $w_\text{avg}= w_\text{max}=4$. 
On the other hand, planar codes usually  have smaller $c$ or larger $w_\text{avg}$ and $w_\text{max}$.

\section{Decoding Performance} \label{sec:Dec}


We assume depolarizing errors. 
%
%
The threshold of a code family and a decoding procedure is estimated by the intersection point of the performance curves  (see \cite{WFSH10} or Fig.~\ref{fig:thrd}).
%

MWPM is the most widely used decoder for 2D topological codes.
The threshold of toric, surface, or XZZX codes using MWPM is  about 15.5\% \cite{WFSH10}, and 
the threshold of color codes using MWPM is about 13\% \cite{WFHH10,Del14}. 
The complexity of decoding toric and surface codes by MWPM is $O(N^2)$, but it is  $O(N^3)$ for the XZZX codes~\cite[supplemental material]{TBFB20} and $O(N^4)$ for the color codes \cite{Del14}.

We proposed MBP 
and it achieves thresholds close to 16\% and 17.5\% on the surface and toric codes, respectively, with  complexity $O(N\log\log N)$ \cite{KL21}.
The decoding  procedure is simply the syndrome-based quaternary message passing on the Tanner graph corresponding to the underlying code, 
CSS or non-CSS, regardless of the layout.
Since MBP can handle binary or quaternary messages, we will use MBP$_4$ with a subscript to emphasize that quaternary decoding  is considered here~\cite[Algorithm~1]{KL21}.

\begin{figure}
	\centering \includegraphics[width=0.5\textwidth]{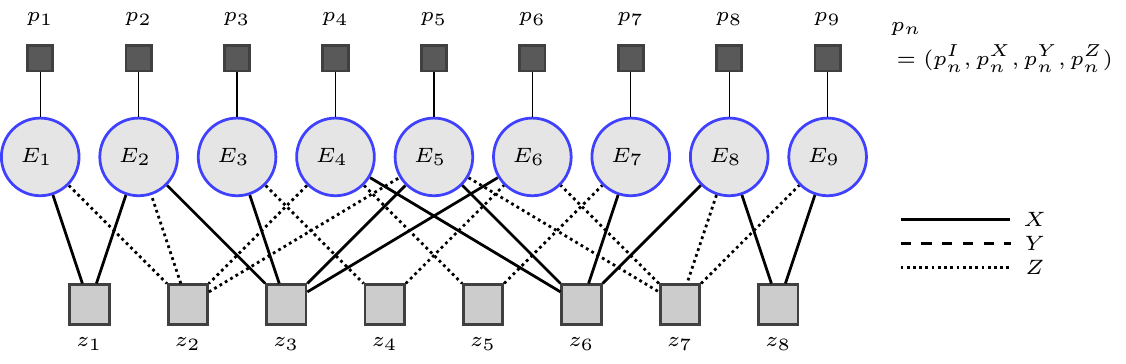}
	\caption{
		 {The Tanner graph of the $[[9,1,3]]$ surface code in Fig.~\ref{fig:surf_3x3_rot}. 
		${E_n\in\{I,X,Y,Z\}}$ is a variable node corresponding to the Pauli error on the $n$-th qubit. $p_n$ is the initial belief of $E_n$. 
		 $z_m\in\{0,1\}$ is the syndrome bit of the $m$-th stabilizer measurement.}
	} \label{fig:surf913_Tanner}		
	\end{figure}

MBP$_4$ is a message passing algorithm on the Tanner graph defined by the check matrix of a code. 
A check matrix $S$ of a code is an $M\times N$ matrix over   $\{I,X,Y,Z\}$, where $M$ is the number of measured stabilizers. 
For example, the $[[9,1,3]]$ surface code in Fig.~\ref{fig:surf_3x3_rot} has
%
\begin{align*}
S = \left[
\begin{smallmatrix}
X &X &I &I &I &I &I &I &I\\
Z &Z &I &Z &Z &I &I &I &I\\
I &X &X &I &X &X &I &I &I\\
I &I &Z &I &I &Z &I &I &I\\
I &I &I &Z &I &I &Z &I &I\\
I &I &I &X &X &I &X &X &I\\
I &I &I &I &Z &Z &I &Z &Z\\
I &I &I &I &I &I &I &X &X\\
\end{smallmatrix} \right]
\end{align*}
and the corresponding Tanner graph is shown in Fig.~\ref{fig:surf913_Tanner}, drawn as a factor graph to also show the initial distribution $p_n$.

MBP$_4$ uses a message normalization parameter $\alpha$.
One can optimize the decoding performance  over $\alpha\in \{1, 0.99, \dots, 0.5\}$ to select an optimum $\alpha^*$,
and this is referred to as adaptive MBP$_4$ (AMBP$_4$) \cite[Algorithm~2]{KL21}.

\begin{figure}
	\centering \includegraphics[width=0.5\textwidth]{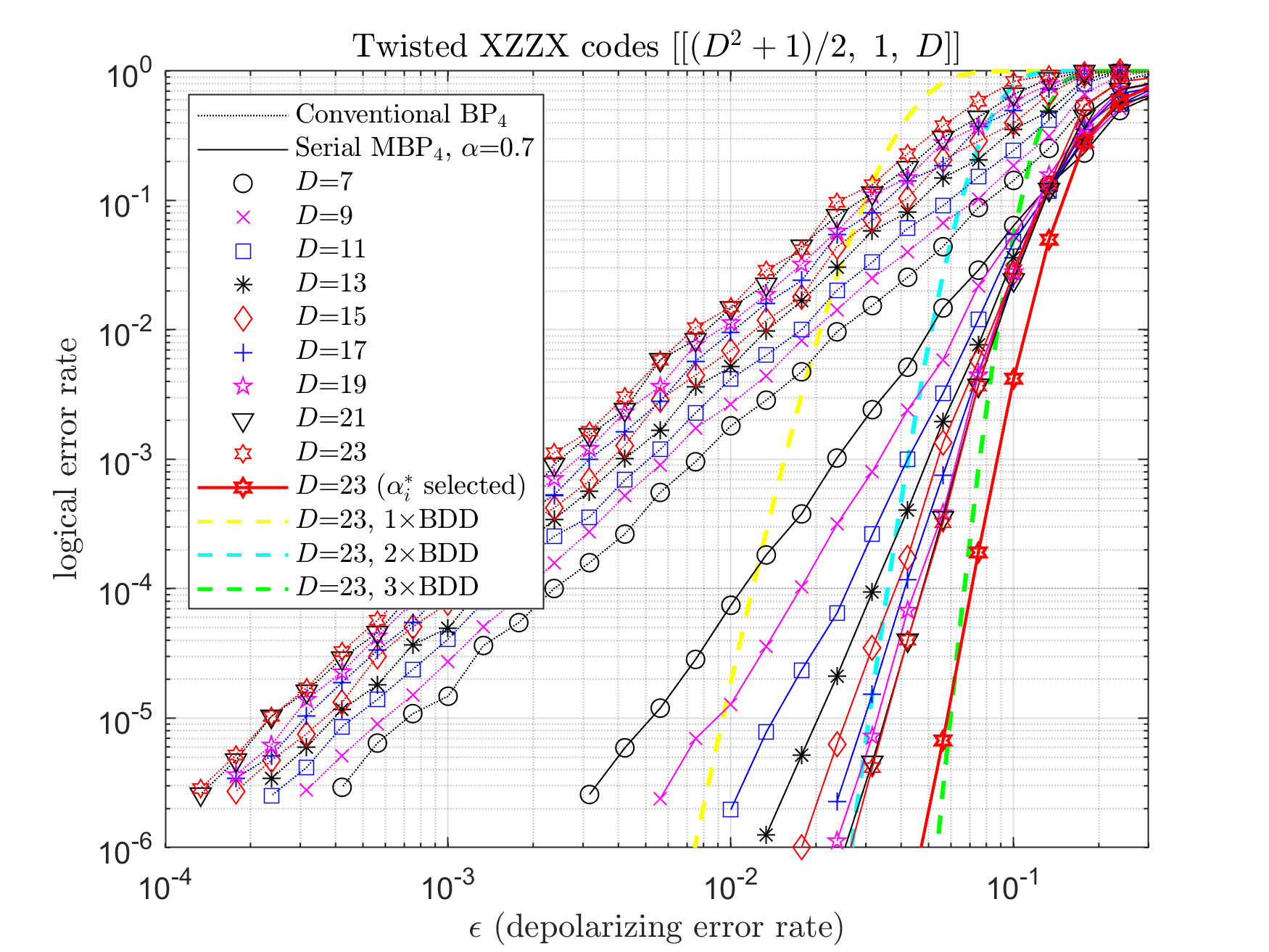}
	\caption{
	MBP decoding performance. Dotted lines are conventional BP$_4$.
	Solid lines are MBP$_4$ with $\alpha=0.7$, which has performance saturation effect because a fixed $\alpha$ is used.
	The bold line is AMBP$_4$ (which uses MBP$_4$ but chooses an optimum $\alpha^*$ and has higher complexity).
	Several bounded-distance-decoding (BDD) performance curves are also shown for reference.
	} \label{fig:BP_a70}		
	\end{figure}

\begin{figure}
	\centering \includegraphics[width=0.44\textwidth]{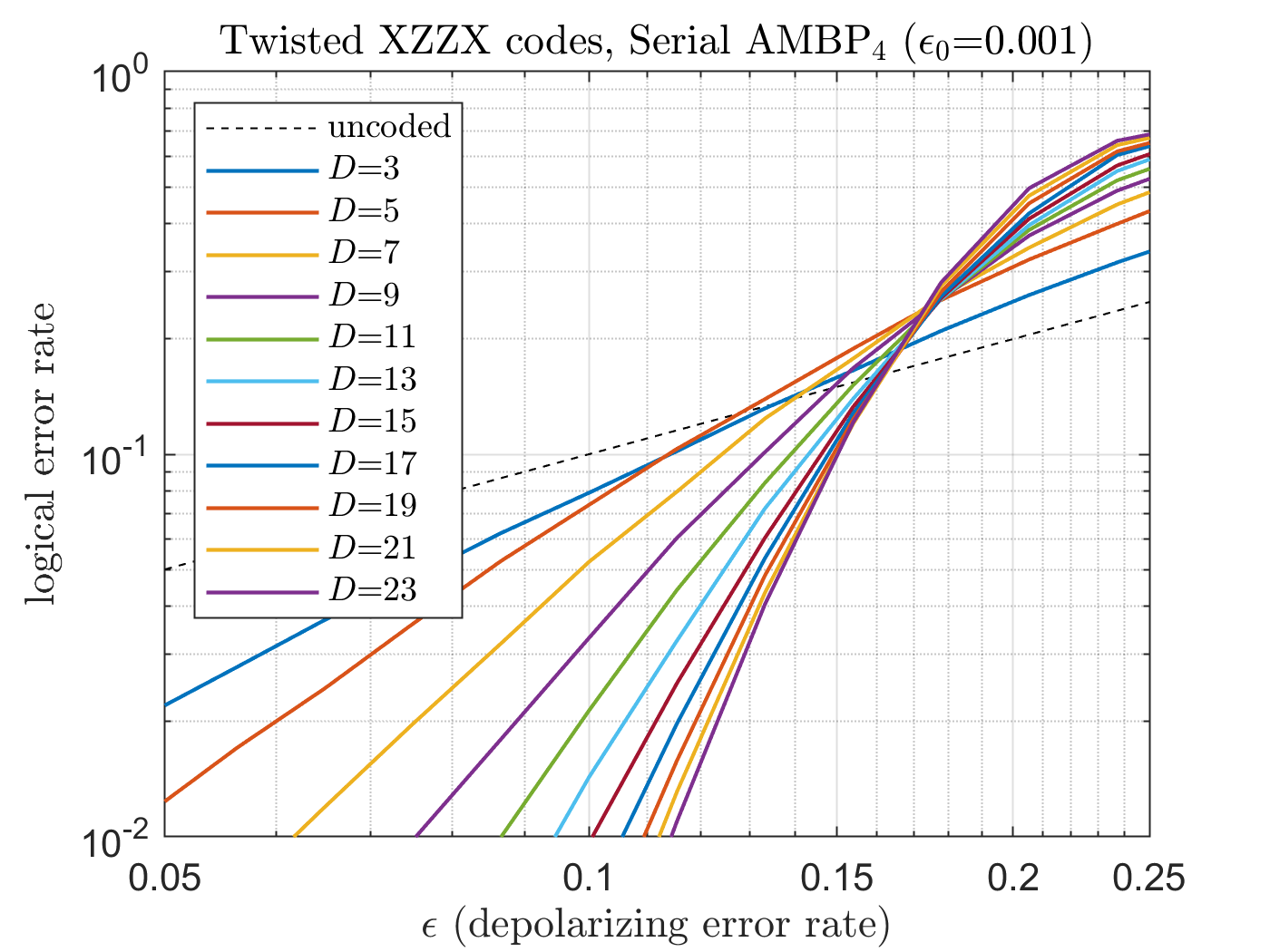}
	\caption{
	The   performance curves of  the $[[(D^2+1)/2,1,D]]$ twisted XZZX codes 
	over depolarizing errors by AMBP$_4$. 
	The intersection point of the performance curves roughly suggests a threshold of 17.5\%.
	} \label{fig:thrd}		
	\end{figure}

First we consider the twisted XZZX codes with parameters $[[(D^2+1)/2,1,D]]$. 
%
In Fig.~\ref{fig:BP_a70}, we show that MBP$_4$ improves   the conventional BP$_4$  with  $\alpha=0.7$.
 {(We collect 100 logical errors for each data point for a figure at this scale.)}
 The performance of AMBP$_4$ for $D=23$ is also shown in Fig.~\ref{fig:BP_a70}.

We apply AMBP$_4$ for different $D$ and show in Fig.~\ref{fig:thrd} that the threshold of AMBP$_4$ on the twisted XZZX codes is close to 17.5\%.
(We collect 10000 logical errors for each data point for a figure at this scale.)
To prevent any performance fluctuation, we use the technique of initializing $p_n$ by a fixed depolarizing rate $\epsilon_0$, 
i.e., fixing $p_n=(1-\epsilon_0,\frac{\epsilon_0}{3},\frac{\epsilon_0}{3},\frac{\epsilon_0}{3})$ regardless of the actual $\epsilon$. 
(See \cite{KL21} for more discussions of this technique).

\begin{figure}
	\centering \includegraphics[width=0.44\textwidth]{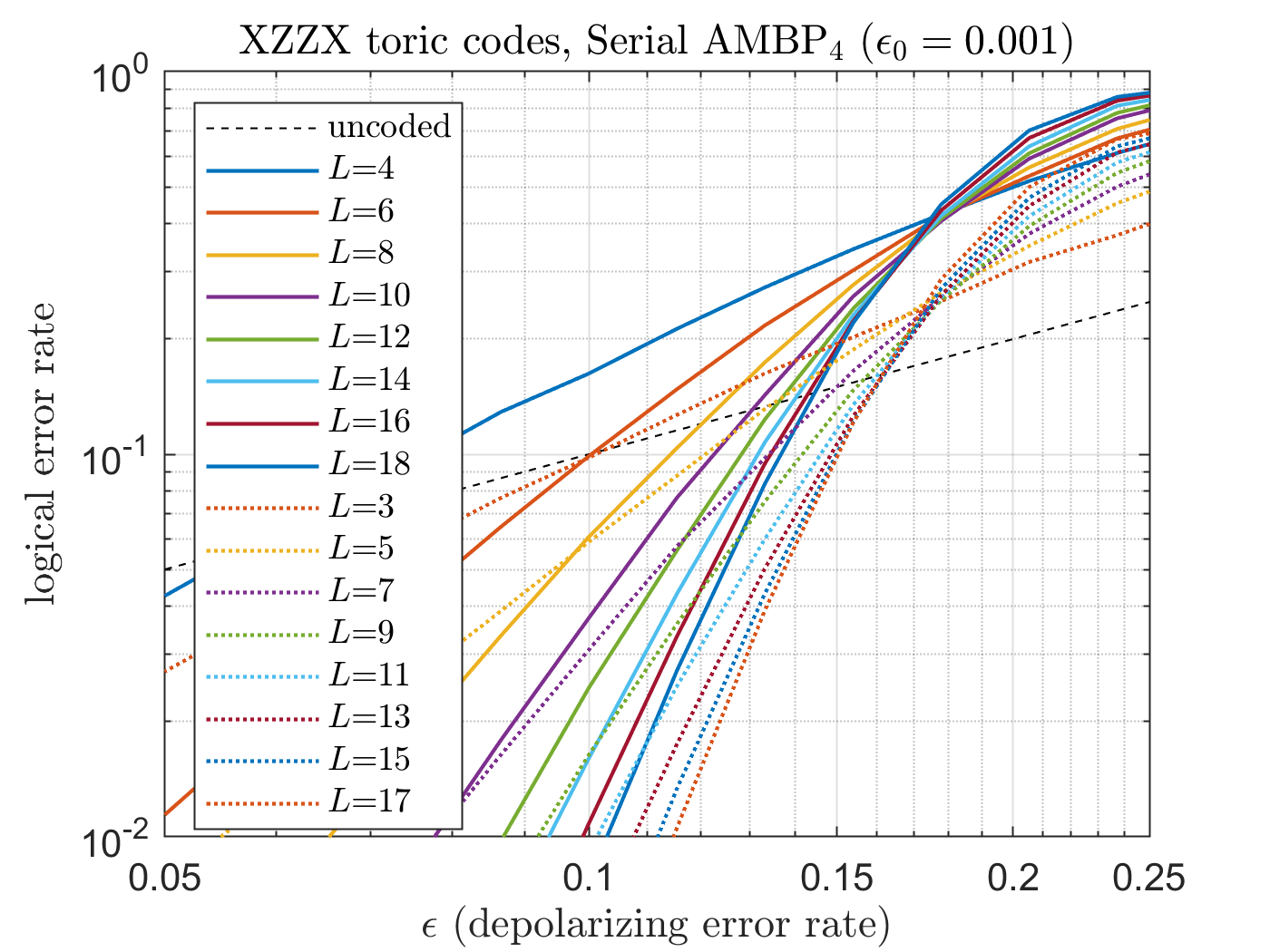}
	\caption{
	 {The decoding performance curves of the  $[[L^2, 2-(L\%2), L]]$ XZZX toric codes over depolarizing errors by AMBP$_4$.
		Both even~$L$ (with efficiency $c=2$) and odd $L$ (with $c=1$) roughly achieve a threshold of 17.5\%.}
	} \label{fig:thrd_nontwist}		
	\end{figure}

 {Denote the error distribution by $p=(p^I,p^X,p^Y,p^Z)$.
	XZZX codes are more effective for biased Pauli errors \cite{ATBFB21},
but they need a tailored MWPM with complexity $O(N^3)$ due to the non-CSS plaquette \cite[supplemental material]{TBFB20}.
For MBP, any distribution $p$ can be supported and the decoding complexity remains the same.
Herein, we focus on   depolarizing errors   for comparison.
The performance curves of AMBP$_4$ on the XZZX toric codes defined in Fig.~\ref{fig:nonCSS_J=0} is shown in Fig.~\ref{fig:thrd_nontwist}.
Both even~$L$ and odd~$L$ roughly achieve a threshold of 17.5\%.}

\begin{figure}
	\centering \includegraphics[width=0.44\textwidth]{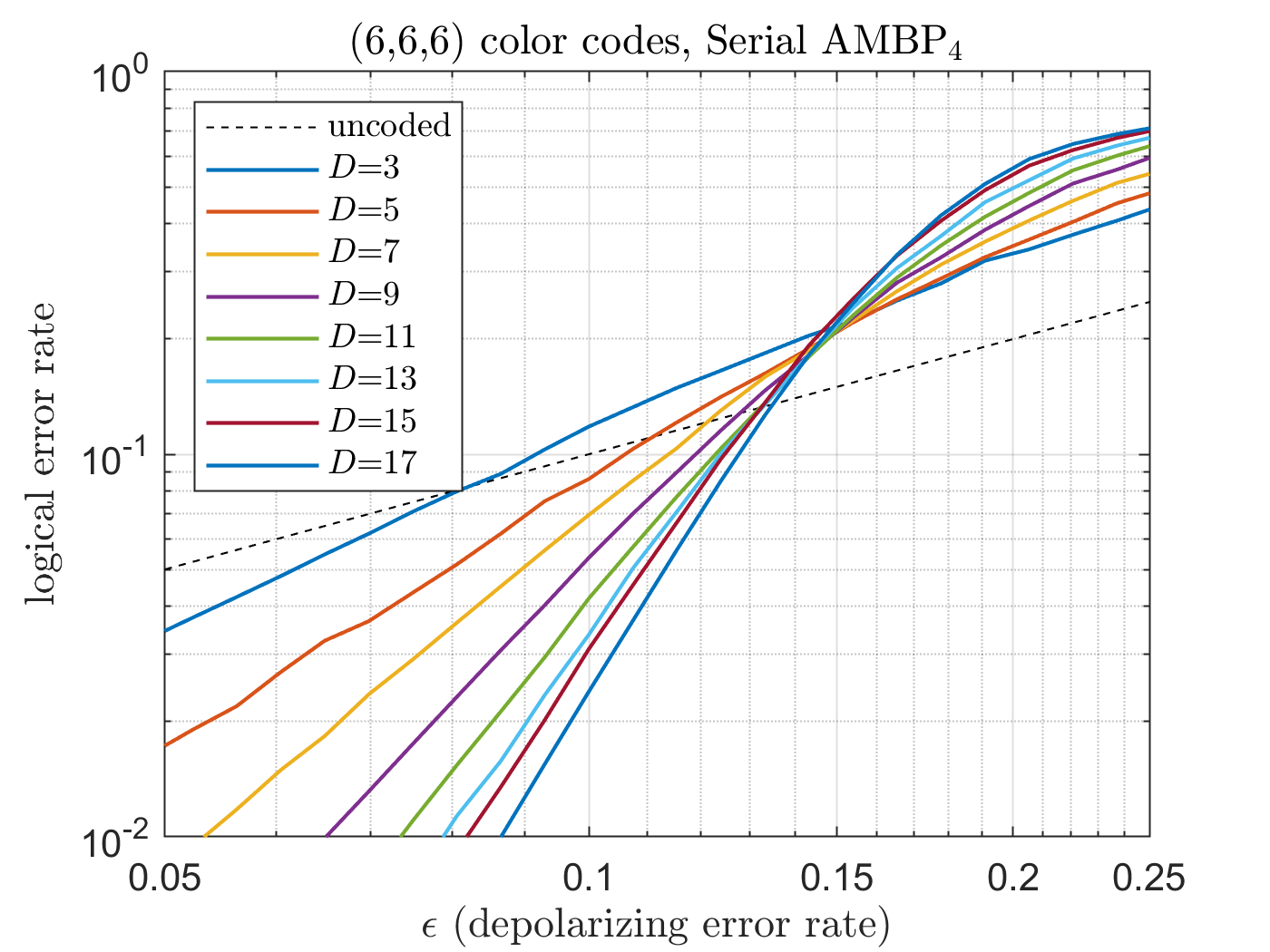}
	\caption{
		 {The   performance curves of the (6,6,6) color codes over depolarizing errors by AMBP$_4$,
		which roughly suggest a threshold of 14.5\%. (For $D=17$, $p_n$ is initialized by a fixed $\epsilon_0=0.042$ to prevent the curve fluctuation.)}
	} \label{fig:thrd_color}		
	\end{figure}

  {For color codes, MWPM needs additional processes since a stabilizer plaquette may have weight higher than four.} 
	MWPM achieves a threshold of 13.3\% on the (4,8,8) color codes without specifying the complexity \cite{WFHH10}
	and a threshold of 13.05\% on the (6,6,6) color codes with complexity $O(N^4)$ \cite{Del14}.
	(Decoding the (4,8,8) color codes is considered relatively harder from the trellis complexity of the code \cite{SAB21}.)
	AMBP$_4$, on the other hand, can decode a color code by just giving its check matrix, without additional processes.
	For comparison in terms of complexity, 
	we decode the (6,6,6) color codes by AMBP$_4$ and the performance curves roughly suggest a threshold of 14.5\%, as shown in Fig.~\ref{fig:thrd_color}.

We summarize all the results  in Table~\ref{tbl:thrd}.

\begin{table}
	\caption{ 			
		Thresholds and complexities of various quantum codes with MWPM- or BP-based decoders over depolarizing errors.
	} 					
	\label{tbl:thrd} 
	{\centering
	$\begin{array}{|l|ll|l|}
	\hline
 	 		&  \text{MWPM \cite{Edm65}} &	& \text{AMBP$_4$ \cite{KL21}}	 \\
 	\text{code family}\ &\text{threshold}	& \text{complexity}  &	\text{threshold}   \\
	\hline                                              
	\text{surface}		& 15.5\%\text{ \cite{WFSH10}}	& O(N^2)\text{ \cite{FWH12}}	& 16\% \\
	\text{toric}		& 15.5\%\text{ \cite{WFSH10}}	& O(N^2)\text{ \cite{FWH12}}	& 17.5\% \\
	\text{color} 		& 13.05\%^\dagger 				& O(N^4)\text{ \cite{Del14}}	&  {14.5\%\text{ (Fig.~\ref{fig:thrd_color})}}\\
    \text{XZZX toric}	& 15.5\%\text{ \cite{ATBFB21}}	& O(N^3)\text{ \cite{TBFB20}}	&  {17.5\%\text{ (Fig.~\ref{fig:thrd_nontwist})}} \\
	\text{XZZX twisted}	&								&								&  {17.5\%\text{ (Fig.~\ref{fig:thrd})}} \\
	\hline
\end{array}$}
\\
$^\dag$: The 13.05\% is rescaled from the 8.7\% for the independent $X$--$Z$ channel by a factor of 3/2 \cite{Del14}.\\
The complexity of AMBP$_4$ is $O(N\log\log N)$ in each case, by a similar analysis as in \cite{KL21}.
\end{table}

\begin{figure}
	\centering \includegraphics[width=0.5\textwidth]{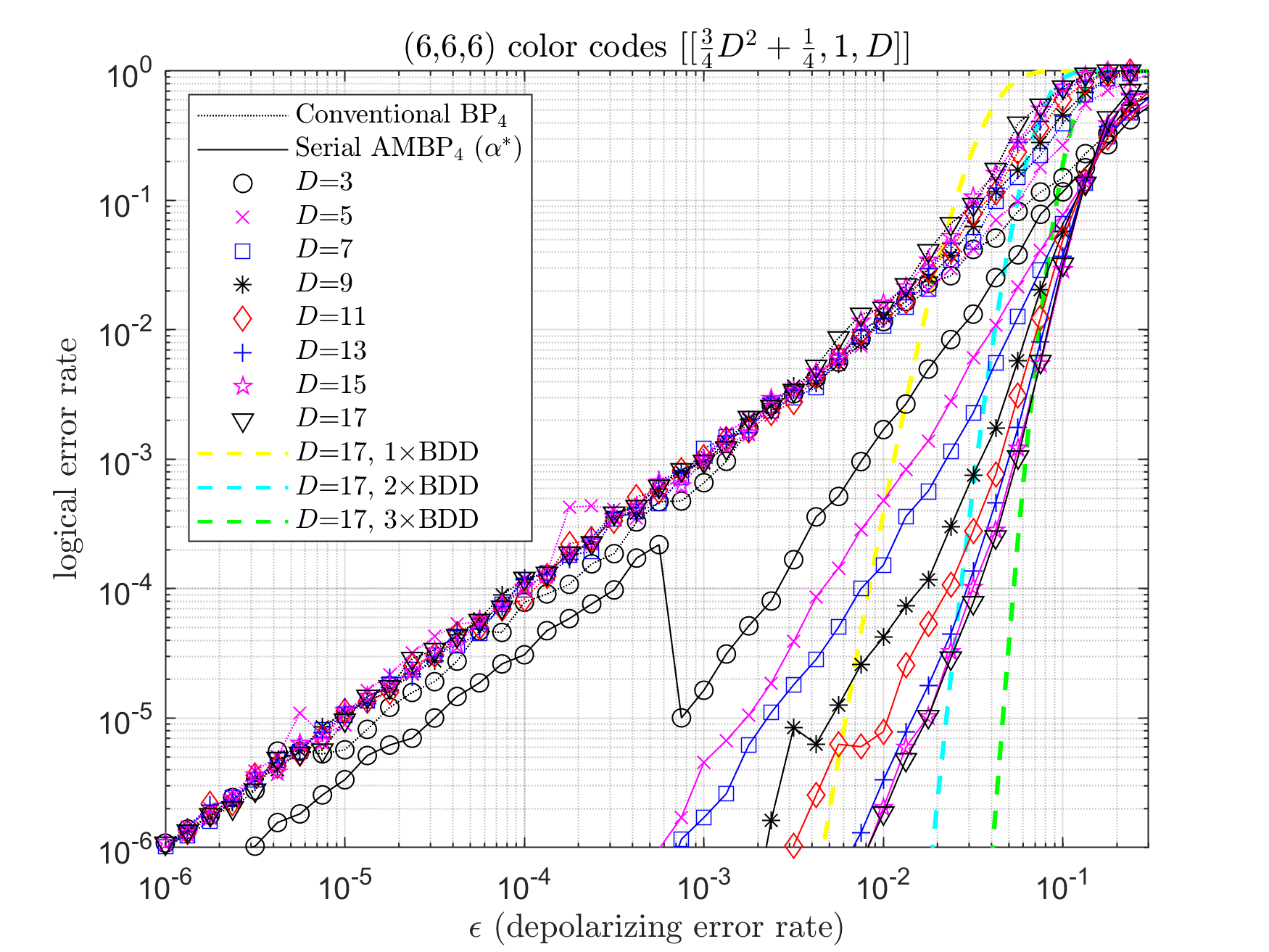}
	\caption{
		 {The   performance curves of the (6,6,6) color codes over depolarizing errors by AMBP$_4$.
		(The curve fluctuation can be prevented by using a fixed $\epsilon_0$ to initialize $p_n$, but this does not improve the error-floor performance.)}
	} \label{fig:color_hex}		
	\end{figure}

We remark that   AMBP$_4$ on color codes is found to have some error floor in performance, as shown in Fig.~\ref{fig:color_hex}.
Since BP is an approximation decoder,  it may encounter this issue for some codes.
In the fault-tolerant case using MWPM on color codes, it seems to have an error floor as well \cite[Fig.~14]{WFHH10}.

\section{Conclusion and Discussions} \label{sec:Conclu}
We compared various 2D topological codes (Table~\ref{tbl:2D_codes}) and their decoding performances by MWPM and MBP (Table~\ref{tbl:thrd}).
We conclude that MBP is easier to adapt to different layouts and tends to have better performance and lower complexity.

It seems that the physical error rate of the intersection point tends to reduce if we keep increasing $D$.
A technique that may prevent this reduction is renormalization group (RG) \cite{DP10} but it needs to concern the lattice layouts. 
Another technique that may improve the threshold value is discussed in Appendix.

MWPM has been extended to  handle measurement and gate errors in FTQC \cite{DKLP02,RHG06,WFSH10,FWH12}.
BP can be extended to correct data and measurement errors simultaneously~\cite{KCL21}.
It is interesting to consider gate errors in BP as well.

\begin{figure}
	\centering \includegraphics[width=0.5\textwidth]{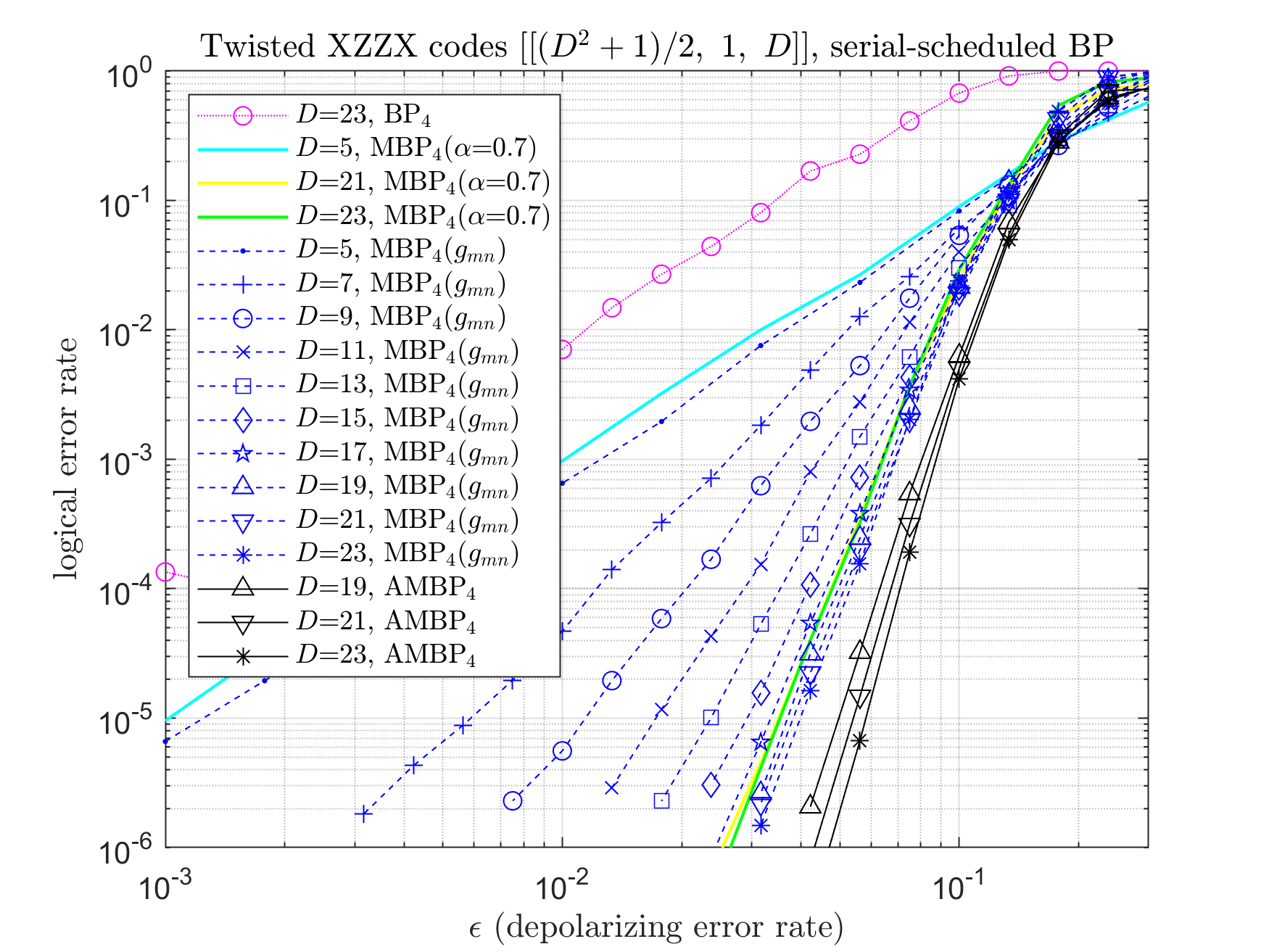}
	\caption{
	MBP decoding performance. 
	The MBP performance saturation effect (Fig.~\ref{fig:BP_a70}) can be improved by using different weight $g_{mn}$ per edge $(m,n)$ of the Tanner graph 
	of the code, where the computation of $g_{mn}$ is referred to \cite[Eq.~(10)]{KL21}.
	AMBP$_4$ still has better performance (but higher complexity since the optimum $\alpha^*$ is determined by many instances of the decoder). 
	} \label{fig:BP_gmn}		
	\end{figure}

\appendix
 Since BP can be seen as a recurrent neural network (RNN), for MBP$_4$, if the weight of each edge ($g_{mn}$) can be determined per iteration \cite[Eq.~(10)]{KL21}, then the decoder will perform better than MBP$_4$ with a fixed $\alpha$, as shown in Fig.~\ref{fig:BP_gmn}.
	Note that AMBP$_4$ still has better performance, which means that
	the thresholds of AMBP$_4$ shown in this paper can be further improved by determining optimum $g^*_{mn}$ per edge per iteration (which may be possibly done by pre-training).

%


\bibliographystyle{IEEEtran}
\bibliography{References}


\end{document}